\documentclass[RNAAS]{aastex62}
\usepackage{amsmath}
\usepackage{mathtools}

\begin{document}

\title{Extended Self-similar Solution for Circumstellar Material-Supernova Ejecta Interaction}
\author{Brighten Jiang}
\affiliation{Lawton Chiles High School, 7200 Lawton Chiles Lane, Tallahassee, FL 32312, USA}
\author[0000-0001-6395-9209]{Shuai Jiang}
\affiliation{Division of Applied Mathematics, Brown University, 182 George Street, Providence, RI 02912, USA}
\author[0000-0002-5814-4061]{V. Ashley Villar}
\affiliation{Center for Astrophysics \textbar{} Harvard \& Smithsonian, 60 Garden Street, Cambridge, MA 02138-1516, USA}

\section{}

In this note, we present a detailed self-similar solution to the interaction of a uniformly expanding gas and a stationary ambient medium, with an application to supernovae interacting with preexisting circumstellar media (Type IIn SNe). This solution was originally presented in \cite{chevalier1982self} with a limited set of solution parameters publicly available. A power-law distribution is assumed for both the expanding supernova ($\rho\propto r^{-n}$) and the stationary circumstellar material ($\rho\propto r^{-s}$).
\cite{chevalier1982self} presented the solution for the case of $s=0$ (a shell-like CSM profile) and $s=2$ (a wind-like CSM profile). In this note, we generalize the solution to $0 \le s \le 2.$ We implement the generalized solution into the Modular Open Source Fitter for Transients ({\tt MOSFiT}; \citealt{guillochon2018mosfit}), an open-source Python package for fitting extragalactic transient light curves.

\section{Self-similar solutions}
We assume the expanding ejecta has a density profile ($\rho$) described by:
\begin{equation}
    \rho(t,r) = t^{-3}(r/tg)^{-n}
\end{equation}
where $g$ is a constant. The CSM is stationary with a density profile described by $\rho(r) = qr^{-s}$ where $q$ is a constant. The interaction of the SNe and CSM results in a shocked region composed of shocked ejecta and shocked ambient gas, separated by a contact discontinuity at radius $R_c$:
\begin{equation}
    R_c = \left(\frac{Ag^n}{q}\right)^\frac{1}{n-s} t^{\frac{n-3}{n-s}}
\end{equation}
where $A$ is a constant and $\lambda = (n-s)/(n-3)$. 
The resulting forward and reverse shocks have radii $R_1$ and $R_2$ respectively. The values $A$, $R_1$, $R_2$ and $R_c$ are dependent on the profile of both the SN ejecta and CSM, and they directly affect the broadband optical light curve of the resulting SN (e.g., \citealt{chatzopoulos2013analytical,villar2017theoretical}). 
As such, our aim is to solve for $A$, $R_1/R_c$, and $R_2/R_c$. 

Recall the standard radial hydrodynamic equations \citep{parker1963interplanetary}:
\begin{align}
\begin{split}
    \frac{\partial u}{\partial t} + u \frac{\partial u}{\partial r} &= -\frac{1}{\rho} \frac{\partial p}{\partial r} \\ 
    \frac{\partial \rho}{\partial t} + \frac{1}{r^2} \frac{\partial (r^2 \rho u)}{\partial r} &= 0 \\
    \frac{\partial p}{\partial t} + u \frac{\partial p}{\partial r} &= \gamma \frac{p}{\rho} \frac{d\rho}{dt}
\end{split}
\end{align}
where $u$ is the radial velocity, $\rho$ density, $p$ pressure. To solve for the quantity $R_2/R_c$, we employ the similarity transform presented in \citet{chevalier1982self}:
\begin{align}
\begin{split}
    \eta &= t ^{-1} r ^\lambda\\\label{eq:4}
    p &= t ^{n-5} r^{2-n} P(\eta)\\
    u &= \frac{r}{t} U(\eta)\\
    \rho &= t ^ {n-3} r^{-n} \Omega(\eta)\\
    C^2(\eta) &= \frac{\gamma P(\eta)}{\Omega(\eta)}
\end{split}
\end{align}
The above similarity transform results in the following systems: 
\begin{align}
\begin{split}
    U^2 - U + (\lambda U - 1)\eta \frac{d U}{d \eta} &= -\frac{(2-n)}{\gamma} C ^2 - \frac{\lambda \eta}{\gamma}\frac{C ^2}{P}\frac{d P}{d \eta } \\
    (n-3)(1-U) + \eta(\lambda U -1)\left(\frac{1}{P}\frac{d P}{d \eta} - \frac{2}{C} \frac{d C}{d \eta}\right) + \lambda \eta \frac{d U}{d \eta} &= 0\\
    P[(n-5) - \gamma(n-3) - U(n-2-n\gamma)] + \frac{d P}{d \eta}\eta (\lambda U -1)(1-\gamma) &= - \frac{2\eta\gamma(\lambda U-1)}{C}P \frac{d C}{d \eta} 
\end{split}
\end{align}
The initial conditions of the inner region at radius $R_2$ are given by:
\begin{align}
\label{eqn:initial-conditions-inner}
\begin{split}
    U_2 &= \frac{1}{4}\Big(\frac{3}{\lambda}+1\Big)\\
    P_2 &= \frac{3}{4}g^n \Big(1-\frac{1}{\lambda}\Big)^2 \\
    C_2 &= \frac{\sqrt{3 \gamma}}{4} \left(1 - \frac{1}{\lambda}\right).
    \end{split}
\end{align}
Furthermore, at the contact discontinuity $R_c$, $U_c = \lambda^{-1}$.
Since $g^n$ is a scaling constant, we set it to 1. We use {\tt scipy}'s \texttt{solve\_ivp} function  with initial conditions from Equations~\ref{eqn:initial-conditions-inner}, and domain $\eta \in (1, \eta_c)$ where $\eta_c$ corresponds to the terminal condition $U(\eta_c) = \lambda^{-1}$. Note that for a fixed $t$:
\begin{align}
    \frac{R_2}{R_c} = \frac{\eta_{2}^{1/\lambda}}{\eta_c^{1/\lambda}},
\end{align}
hence $R_2/R_c = \eta_c^{-1/\lambda}$ as we assumed that $\eta_2 = 1$.


For the quantity $R_1/R_c$, we employ the similarity transform from \cite{parker1963interplanetary}:
\begin{align}
\begin{split}
    \eta &= t r^{-\lambda}\\\
    u &= \frac{r}{t} U(\eta)\\ 
    \rho &= \Omega(\eta) r ^{-s}\\
    p &= r ^{2-s}  t^{-2}  P(\eta)\\
    C^2(\eta) &= \frac{\gamma P(\eta)}{\Omega(\eta)}\\
\end{split}
\end{align}
The transformation gives the following system for the outer shock region from $R_1$ to $R_c$: 
\begin{align}
\begin{split}
    [(1-\lambda U) ^2 - \lambda ^2C^2]\eta \frac{dU}{d\eta} &= U(1-U)(1-\lambda U) + \frac{C^2}{\gamma}[(2\lambda - 2+ s) - 3\gamma\lambda U]\\
    2[(1-\lambda U) ^2 - \lambda ^2C^2]\frac{\eta}{C}\frac{dC}{d\eta}&= 2 + U[1-3\lambda-3\gamma +\lambda \gamma] + 2\gamma \lambda U^2 + \frac{C^2}{1-\lambda U}[(-\frac{2\lambda ^2}{\gamma}- \frac{s\lambda }{\gamma} - 2\lambda + s \lambda + \frac{2 \lambda}{\gamma}) +2 \lambda ^2 U] \\
    [(1-\lambda U) ^2 - \lambda ^2C^2]\frac{\eta}{P} \frac{dP}{d\eta} &= 2 + U(s-2-2\lambda  + \lambda \gamma - 3\gamma) + \lambda U^2(2-s+2\gamma) + \lambda C^2(s-2) 
\end{split}
\end{align}
with initial conditions at $R_1$
\begin{align}
\begin{split}
    U_1 &= \frac{2}{\lambda(\gamma +1)}\\
    P_1 &= 2\rho_c\frac{2}{\lambda^2 (\gamma+1)}\\
    \Omega_1 &= \rho_c\frac{\gamma +1}{\gamma -1}
\end{split}
\end{align}
where $\rho_c$ is a scaling constant. 
Using $U_c = \lambda^{-1}$ as before, the calculation follows analogously to the calculation of $R_2/R_c$ from above. 


The two solutions for the inner and outer region can be used to find $A$, the scaling constant.
Let $P_c/P_1$ be the pressure at the contact discontinuity divided by the initial pressure value for the outer region. 
Let $P_c/P_2$ be the analogous ratio for the inner region. The constant $A$ can then be defined by \citet{chevalier1982self}:
\begin{align}
    A = \left(\frac{P_c}{P_1}\right)^{-1} \left(\frac{P_c}{P_2}\right) \left(\frac{3-s}{n-3}\right)^2.
\end{align}

\section{Application to Type IIn Supernova Light Curves}

\begin{figure}[h!]
\begin{center}
\includegraphics[scale=0.65,angle=0]{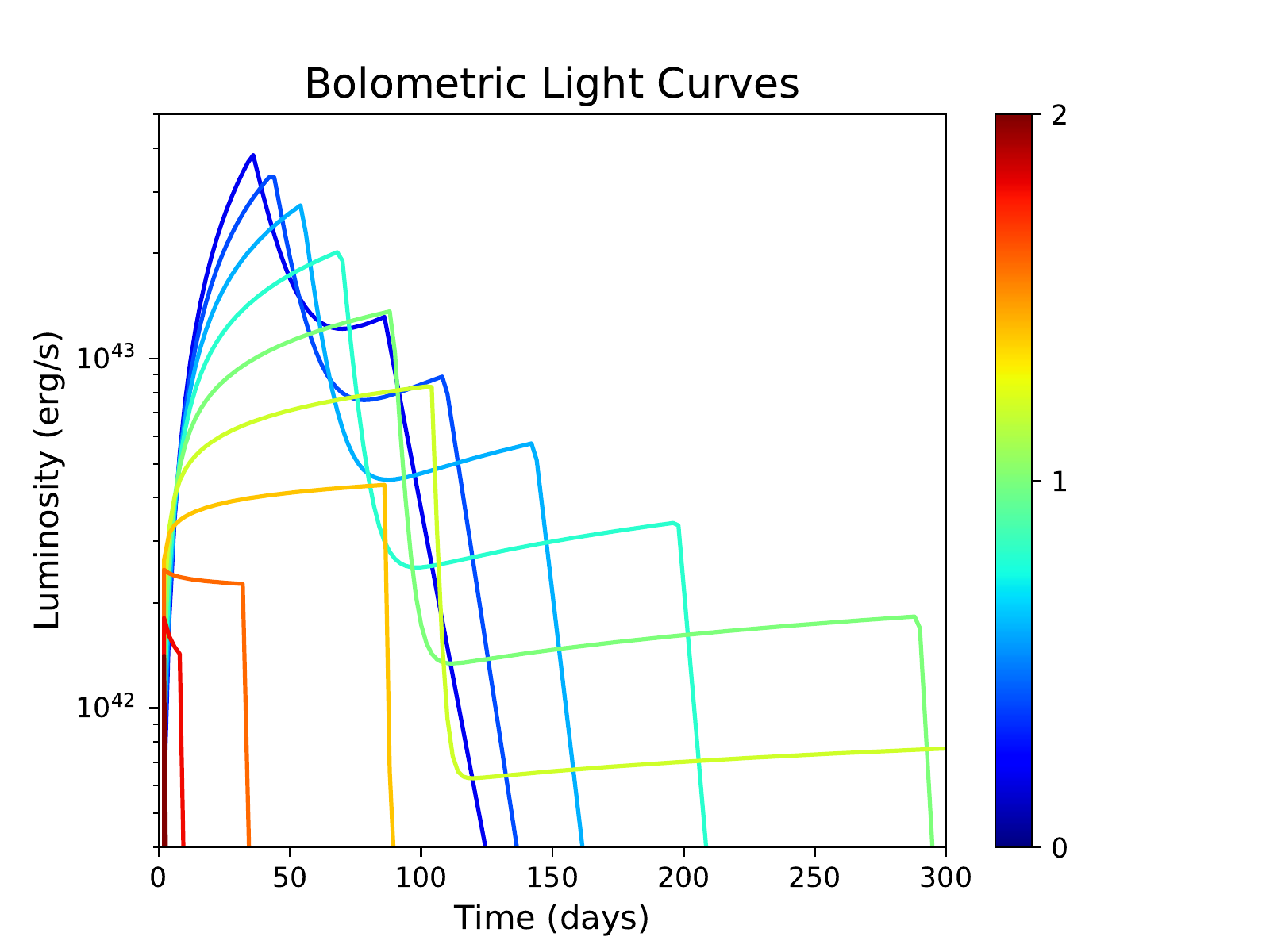}
\caption{Type IIn SN model varying the CSM shape from a shell-like profile ($s=0$, blue) to a wind-like profile ($s=2$, red). We set the other model parameters to the following:  SN ejecta mass $M_\mathrm{ej} = 12 M_\odot$, SN ejecta velocity $v_\mathrm{ej}=5000$ km s$^{-1}$, ejecta index $n=12$, CSM mass $M_\mathrm{CSM} = 1 M_\odot$ , CSM inner radius $R_0=10^{14}$ cm, CSM inner density $\rho_{CSM}(R_0)=10^{-13}$ and CSM/ejecta opacity $\kappa=0.34$ cm$^2$ g$^{-1}$. \label{fig:f1}}
\end{center}
\end{figure}

Type IIn SNe are a class of core-collapse SNe with characteristic narrow hydrogen emission during the photospheric phase \citep{filippenko1997optical}. In the canonical model, these Type IIn SNe arise from the collapse of massive stars in dense CSM environments originating from enhanced mass-loss from the progenitor. Many models (eg., \citealt{smith2007shell,chatzopoulos2013analytical,moriya2013analytic,ofek2014interaction}) assume that the SN luminosity is dominated by shocks formed in the interaction between the stationary CSM and the high-velocity SN ejecta. In the simple, one-zone model explored here, the luminosity of the SN results from the conversion of both the forward and reverse shocks' kinetic energy into heating: 

\begin{align}
    L = \epsilon\frac{d}{dt}(\frac{1}{2}M_\mathrm{sw}v_\mathrm{sh}^2)= \epsilon M_\mathrm{sw}v_\mathrm{sh} \frac{dv_\mathrm{sh}}{dt} + \epsilon\frac{1}{2}\frac{dM_\mathrm{sw}}{dt}v_\mathrm{sh}^2
\end{align}

, where $M_\mathrm{sw}$ is the mass swept up by the shock and $v_\mathrm{sh}$ is the shock velocity. Both the luminosity and diffusion timescale depend heavily on $R_1/R_c$, $R_2/R_c$ and $A$, implying that they can be directly measured from the broadband optical light curves. In Figure \ref{fig:f1}, we demonstrate the effect of $s$, the CSM profile, on the light curve. 

We incorporate our extended solutions to these parameters into the open-source code {\tt MOSFiT}. We additionally publish a Jupyter notebook highlighting our results on GitHub\footnote{\url{https://github.com/Brightenj11/SSS-CSM}}.


\acknowledgments

VAV is supported by a Ford Foundation Dissertation Fellowship.

\software{matplotlib \citep{Hunter:2007}, numpy \citep{numpy}, scipy \citep{2019arXiv190710121V}}

\bibliography{mybib}

\end{document}